\documentstyle[twoside,11pt]{article}
\newcommand{\ra}{\rightarrow}

\newcommand{\vp}{\varphi}
\newcommand{\vt}{\vartheta}

\normalsize
\begin{document}
\bibliographystyle{plain}
%\tableofcontents
%\makeindex

\begin{titlepage}
\begin{flushright}
ESI-1993-18\\
March 1993
\end{flushright}
\vfill
\begin{center}
{\Large \bf Saddle points of stringy action} \\[40pt]
P. Bizon* \\
Erwin Schr\"odinger Institut f\"ur Mathematische Physik \\
 Wien, Austria
\vfill
{\bf Abstract} \\
\end{center}
It is shown that Einstein-Yang-Mills-dilaton theory has a countable
family of static globally regular solutions which are purely magnetic
but uncharged. The discrete spectrum of masses of these solutions is
bounded from above by the mass of extremal Gibbons-Maeda solution.
As follows from linear stability analysis all solutions are unstable.

\vfill
\noindent *) On leave of absence from Institute of Physics, Jagellonian
University, Cracow, Poland. E-mail address: bizon@awirap.bitnet
\end{titlepage}

\section{Introduction}
In a seminal paper [1] Bartnik and Mckinnon (BM) have discovered a countable
family of globally regular static spherically symmetric
solutions of
the Einstein-Yang-Mills (EYM) equations. A rigorous
proof of existence of BM solutions was given by Smoller et al. [2].
Recently, Sudarsky and Wald have proposed a heuristic argument [3],
in the spirit of Morse theory, which
explains existence and properties of BM solutions. This
argument exploits the existence of topologically
nonequivalent multiple
vacua in the $SU(2)$-YM theory (which is related to the fact that
$\pi_3(SU(2)) \simeq Z$) and is essentially insensitive to the concrete
form of the coupling,
 which suggests
that there should exist solutions similar to BM solutions in other
theories involving the $SU(2)$-YM field. Indeed, such solutions were found in
YM-dilaton theory [4,5] and remarkable parallels between these solutions
and BM solutions were observed [5].

The YM-dilaton theory and the Einstein-YM theory may
be embedded in a single Einstein-YM-dilaton theory governed by the action
$$
S = \int d^4 x \sqrt{-g} \left[\frac{1}{G} R -2 (\nabla \phi)^2
- e^{-2a \phi} {\cal F}^2 \right] \;, \eqno(1)
$$
where $R$ is a scalar curvature, $\phi$ is a dilaton, $\cal F$ is a
Yang-Mills curvature, $G$ is Newton's constant
and $a$ is the dilaton coupling constant.
This theory is characterized by a dimensionless parameter
$\alpha=G/a^2$.
When $\alpha=0$ the action (1) reduces to the YM-dilaton
theory. When $\alpha \rightarrow \infty$ the action (1) becomes the
Einstein-YM theory (plus trivial kinetic term for the scalar field).
 Finally, the case $\alpha=1$ corresponds to
the low-energy action of heterotic string theory.

The aim of this paper is to show that the theory defined by the
action (1) has (for all values of $\alpha$) static
spherically symmetric globally
regular solutions with the following properties:
\begin{enumerate}
\item[a)] there exist a countable family of solutions ${X_n}$
($n\in N$),
\item[b)] the total mass $M_n$ increases with $n$ and is bounded
from above,
\item[c)] the solution $X_n$ has exactly $n$ unstable modes.
\end{enumerate}
The solutions $X_n$ depend continously on $\alpha$ and
interpolate smoothly between the YM-dilaton solutions ($\alpha=0$)
and BM solutions ($\alpha=\infty$). In the limit $n \ra \infty$ the
solution $X_n(\alpha)$ tends to the abelian extremal charged dilatonic black
hole [6,7], whose mass therefore provides un upper bound for the spectrum
$M_n$. These results were previously announced by the author in [5] and
later the case $\alpha=1$ was analysed in [8].

The paper is organized as follows. In the next Section the field
equations are derived and some scaling properties are discussed.
In Section 3 the explicit abelian solutions are described.
In Section 4 the numerical non-abelian solutions are presented and
their qualitative properties are discussed. Section 5 is devoted to
linear stability analysis.

\section{Field equations}
We are interested in static spherically symmetric configurations. It is
convenient to parametrize the metric in the following way
$$
ds^2 = - A^2 N dt^2 + N^{-1}dr^2 + r^2(d\vt^2 + \sin^2 \vt d\vp^2)\;, \eqno(2)
$$
where $A$ and $N$ are functions of $r$.

I assume that the electric part of the YM field vanishes (actually
this is not a restriction because one can show, following the argument
given in [9], that there are no globally regular static solutions with nonzero
electric field). The
purely magnetic static spherically symmetric $SU(2)$ YM connection can be
written, in the abelian gauge, as [10]
$$
e{\cal A} = w \tau_1 d\vt + (\cot \vt \tau_3 + w \tau_2) \sin \vt d\vp\;,
\eqno(3)
$$
where $\tau_i$ ($i = 1,2,3)$ are Pauli matrices and $w$ is a function
of $r$.
The corresponding YM curvature ${\cal F} = d{\cal A} + e {\cal A}
\wedge {\cal A}$ is given by
$$
e{\cal F} =  w' \tau_1 dr \wedge d\vt + w' \tau_2 dr \wedge \sin \vt d\vp
 - (1-w^2) \tau_3 d\vt \wedge \sin \vt d\vp\;,
\eqno(4)
$$
where prime denotes derivative with respect to $r$.

For these anz\"{a}tze  and for $\phi=\phi(r)$ the action (1) gives the
lagrangian (where $S=16\pi\int L dt$)
$$
L =  \int_{0}^\infty \; A ( m' - U ) dr\;\;,
\eqno(5)
$$
where the mass function $m(r)$ is defined by $N = 1 - 2Gm/r$ and
$$
U = \frac{1}{2} r^2 N \phi'{}^2  + \frac{1}{e^2} e^{-2 a \phi}
\left[ N w'{}^2 + \frac{(1-w^2)^2}{2r^2} \right]\;,
\eqno(6)
$$
Hereafter, for convenience, I put the coupling constants
$e=a=1$ which means that I choose $a/e$ as the unit
of length and $1/ea$ as the unit of energy. Then the system is
characterized by one dimensionless parameter $\alpha=G/a^2$.
Variation of $L$ with respect of $m,A,w$, and $\phi$ yields the field
equations\footnote{The principle of symmetric criticality
(see, R.Palais, Comm.Math.Phys. {\em 69} (1979) 19) guaranties
that the variation
of $S$ within the spherically symmetric ansatz gives the correct
equations of motion.}
$$
m' = \alpha U\;,
\eqno(7)
$$
$$
A' =  \alpha A ( r \phi'{}^2 + \frac{2}{r} e^{-2 \phi} w'{}^2 )\;,
\eqno(8)
$$
$$
(A N e^{-2 \phi} w')' +  \frac{1}{r^2} A e^{-2 \phi} w(1 - w^2) = 0\; ,
\eqno(9)
$$
$$
(r^2 A N \phi')' +  2 A e^{-2 \phi}
\left[ N w'{}^2 + \frac{(1-w^2)^2}{2r^2} \right] = 0.
\eqno(10)
$$

Note that $L$ has a characteristic for general relativity "pure constraint"
form, namely the integrand in Eq.(5) is the hamiltonian
constraint, Eq.(7), multiplied by the lapse function $A$.
Actually $L$ is not differentiable because the variation of $L$
gives an
unwanted surface term at infinity. To remedy this one has to add
to $L$ the Regge-Teitelboim correction term,
$-A(\infty)m(\infty)$. Then $L'=L-A(\infty)m(\infty)$ has a well-defined
functional derivative (since I deal with globally regular solutions I
assume that $m(0)=0$). It is convenient to define the energy functional
$$
E = -L' = \int_{0}^\infty (AU - A'm) dr\;,
\eqno(11)
$$
which on shell is equal to the total mass $m(\infty)$ (assuming
the boundary condition $A(\infty)=1$ i.e. that $t$ is proper time at
spatial infinity).

Eqs.(7)-(10) have a scaling symmetry: if $w,\phi,N$ and $A$ are solutions
so are
$$
w_{\lambda}(r) = w(e^{\lambda} r) \;, \eqno(12a)
$$
$$
\phi_{\lambda}(r) = \phi(e^{\lambda} r) + \lambda \;, \eqno(12b)
$$
$$
N_{\lambda}(r) = N(e^{\lambda} r) \qquad  \left[
m_{\lambda}(r) = e^{-\lambda} m(e^{\lambda} r) \right]\;, \eqno(12c)
$$
$$
A_{\lambda}(r) = A(e^{\lambda} r).  \eqno(12d)
$$
Under this transformation the energy functional scales
as follows
$$
E_{\lambda} = e^{-\lambda} E \;.              \eqno(13)
$$
The existence of this scaling symmetry does not exclude
globally regular solutions
because for the variation induced by the transformation (12)
      the energy is not extremized since
$\delta \phi(\infty)$ is nonzero.
 Hereafter, I will assume
that all solutions satisfy $\phi(\infty)=0$, which can always be
set by the transformation (12). This choice sets the scale of
energy in the theory.

\section{Abelian solutions}
In the $U(1)$ sector of the theory (i.e. in Einstein-Maxwell-dilaton theory)
there are no static globally
regular solutions but there are known explicit black hole solutions
of Eqs.(7)-(10).
The uncharged solution is Schwarzschild
$$
w = \pm 1 \;, \qquad \phi = 0 \;, \qquad m = M = const \;, \qquad A = 1.
\eqno(14)
$$
Charged black hole solutions were found by Gibbons and Maeda [6] (and
later rediscovered in [7]). In the so called extremal limit the area of
the horizon of these charged dilatonic black holes goes to zero and
the resulting spacetime has a null naked singularity. This
singular extremal solution will play an important role in our discussion
of non-abelian solutions. It has a very simple form in isotropic coordinates
$$
ds^2 = -e^{-2\alpha \phi} dt^2
+ e^{2\alpha \phi} (d\rho^2 + \rho^2 d\vt^2 + \rho^2 \sin^2 \vt d\vp^2)
\eqno(15a)
$$
where
$$
\phi = \frac{1}{1+\alpha} \ln \left( 1 +
\frac{\sqrt{1+\alpha}}{\rho}\right)
\eqno(15b)
$$
and $w=0$, so
the YM curvature is
$$
{\cal F} = - \tau_3 \; d\vartheta \wedge \sin \vartheta d\varphi \;,
\eqno(15c)
$$
which corresponds to the Dirac magnetic monopole with unit magnetic charge
(where the unit of charge is $1/e$).
The total mass is $m(\infty)=1/\sqrt{1+\alpha}$.
For $\alpha=0$ the solution (15) reduces to the dilatonic magnetic monopole
discussed in [5]. For $\alpha=\infty$ the solution (15) reduces to the
extremal Reissner-Nordstrom black hole (before taking the limit $\alpha
\ra \infty$ one has to make rescaling $\rho \ra \sqrt{\alpha}\rho$).

\section{Non-abelian solutions}
In order to construct solutions which are globally regular we have to
impose the boundary conditions which ensure regularity at $r=0$ and
asymptotic flatness.
The asymptotic solutions of Eqs.(7)-(10)
satisfying these requirements are
$$
\pm w = 1 - br^2 + O(r^4) \;, \eqno(16a)
$$
$$
\phi = c - 2 e^{-2c}b^2r^2 + O(r^4) \;, \eqno(16b)
$$
$$
N = 1 - 4 \alpha e^{-2c}b^2 r^2 + O(r^4) \;, \eqno(16c)
$$
$$
A = d \ (1 + 4 \alpha e^{-2c}b^2 r^2) + O(r^4)  \eqno(16d)
$$
near $r=0$, and
$$
\pm w = 1 - \frac{B}{r} + O(\frac{1}{r^2}) \;, \eqno(17a)
$$
$$
\phi = C - \frac{D}{r} + O(\frac{1}{r^4}) \;,  \eqno(17b)
$$
$$
N = 1 - \frac{2\alpha M}{r} + \frac{\alpha D^2}{r^2} + O(\frac{1}{r^3})
\;, \eqno(17c)
$$
$$
A = 1 - \frac{\alpha D^2}{2 r^2} + O(\frac{1}{r^4}) \eqno(17d)
$$
near $r=\infty$.
Here $b,c,d,B,C,D$, and $M$ are arbitrary constants.
All higher order terms in the above
expansions are uniquely determined, through recurrence relations,
by $b,c$ and $d$ in (16), and $B,C,D$ and $M$ in (17).
The constant $D$ is usually referred to as the dilaton charge while $M=
m(\infty)$ is the total mass.

{\em Remark 1.} Notice that, for the asymptotic behaviour (17a), the
radial magnetic curvature, ${\cal B}_r = \tau_3 (1-w^2)/r^2$, falls-off
as $1/r^3$, and therefore all globally regular solutions have
zero YM magnetic charge. In accord with this, in the asymptotic expansion
of $g_{00}=A^2 N$ the $1/r^2$ term vanishes.

{\em Remark 2.}
One can easily show (see [5]) that for global
 solutions satisfying the above boundary conditions,
 a function $w$ oscillates around zero between
$-1$ and $1$, while $\phi$ is monotonically decreasing.

Now, let us assume that near $r=0$ there exist a family of local
solutions defined by the expansion (16). Note that this is a
nontrivial statement because the point $r=0$ is a singular point
of Eqs.(7)-(10), hence the formal power-series
expansion (16) may have, in principle, a zero radius of convergence.
The initial parameters $c$ and $d$ are irrelevant since they can be chosen
arbitrarily by the scaling (12) and by time rescaling, respectively.
Thus,
 effectively we have a one-parameter family parametrized by $b$.
For generic $b$ the solution will not
satisfy the asymptotic conditions (17) (in fact, the solution
may even become singular at some finite distance). The standard numerical
strategy, called the shooting method, is to find an
initial value $b$ for which the local solution extends to a
global solution with the asymptotic behaviour (17). I have found that
 for generic orbits with
$b<b_{\infty}(\alpha)$ a
function $w$ oscillates finite number of times
in the region between $w=-1$ and $w=1$ and then goes to $\pm \infty$.
For $b>b_{\infty}(\alpha)$
all orbits become singular at a
finite distance (in a sense that $w'$ becomes infinite).

The numerical results strongly indicate that for all values
of $\alpha$ there exist a countable
family of initial values $b_n$ ($ n \in N$) determining globally
regular
solutions. Here the index $n$ labels the number
of nodes of the function $w$. In Table 1 are displayed
the initial values and
masses of the first five
solutions for $\alpha=1$. The initial value $c$ is determined by the
condition that the dilaton vanishes at infinity i.e. $C=0$.
The functions $w$, $\phi$ and $N$ are graphed
in Figs.1-3.
% Figs.1-3
\begin{table} [h]
\caption{Initial data $(b,c)$ and masses of the first five globally
regular solutions for $\alpha=1$.}
$$
\begin{tabular}{|c|c|c|c|} \hline
$n$ & $be^{-c}$ & $c$ & $M$ \\ \hline
1 & 0.1666666 & 0.9311 & 0.5773\\
2 & 0.2318001 & 1.7925 & 0.6849\\
3 & 0.246861 & 2.6919 &  0.7035\\
4 & 0.249483 & 3.5974 & 0.7065 \\
5 & 0.249915 & 4.5043 &  0.7070\\  \hline
\end{tabular}
$$
\end{table}
%Figs.4,5

In Figs.4 and 5 are shown the functions $w_1$ and $N_1$ for three
different values of $\alpha$. For all $\alpha$
the solutions display three characteristic regions. The energy density
 is concentrated in the inner core region $r<R_1$, where $R_1$ is
approximately the location of the first zero of $w$.
 In
the second region, $R_1 < r <R_2$, where $R_2$ is approximately the
location of the last but one zero of $w$, the function $w$ slowly
oscillates around $w=0$.
Finally,
in the asymptotic region
 $r>R_2$, the function $w$ goes
monotonically to $w=\pm 1$ (hence the YM magnetic charge is gradually screened)
and for $r \rightarrow \infty$ the solution tends to the Schwarzschild
solution (14).

Let us consider the limit $n \ra \infty$. In this limit $R_1 \ra 0$ and
$R_2 \ra \infty$, so the second
region
covers the whole space. As $n$ grows the amplitude of oscillations of the
function $w$ decreases and goes to zero as $n \rightarrow \infty$ (see Fig.1).
Thus, for $n \rightarrow \infty$ the solution tends (nonuniformly)
to the extremal Gibbons-Maeda solution (15).
This is clear from Figs.2 and 3, where the solution (15) for $\alpha=1$,
expressed in Schwarzschild coordinates
$$
N_{abel} = \left(1 - \frac{1}{1+\sqrt{1+2r^2}} \right)^2 \;, \eqno(18)
$$
and
$$
\phi_{abel} = \frac{1}{2} \ln \left(\frac{1+\sqrt{1+2r^2}}{-1+
\sqrt{1+2r^2}} \right) \;, \eqno(19)
$$
is included for comparison with non-abelian solutions.

For all $n$ and $\alpha$ the metric coefficient
$N$ has one minimum $N_{min}$ located approximately at $R_1$.
For fixed $n$, $N_{min}$ decreases with $\alpha$ (see Fig.5). Also for given
$\alpha$, $N_{min}$ decreases with $n$ (see Fig.3) and
 $N_{min} \rightarrow 1/(1+\alpha)^2$ as $n \ra \infty$.

The total
mass $M_n$ increases with $n$ and for $n \ra \infty$ goes
to $1/\sqrt{1+\alpha}$.
Because our scale of energy breaks down for $\alpha \ra \infty$ (i.e. $a=0$),
it is more convenient to compare masses of solutions with different values
of $\alpha$
using $1/e\sqrt{G+a^2}$ as the unit of energy
instead of $1/ea$. This corresponds to rescaling
$\tilde M=\sqrt{1+\alpha}\;
M$. Then for all $\alpha$ one has $ lim_{n \ra \infty} \tilde M_n = 1$.
It turns out that for given $n$ the mass $\tilde M_n$ is increasing
approximately linearly with $\beta=\alpha/(1+\alpha)$. This is
shown in Fig.6.
%Fig.6
In [5] it was proven in the case $\alpha=0$
that the energy is equal to minus dilaton charge $D$.
This property still holds for $\alpha>0$. To show this I first derive a simple
virial identity. Consider
a one-parameter family of field configurations defined by
$$
w_{\lambda}(r)=w(\lambda r)\;, \qquad \phi_{\lambda}(r)=\phi(\lambda r)\;,
\qquad N_{\lambda}(r)=N(\lambda r)\;, \qquad A_{\lambda}(r)=A(\lambda r)\;.
$$
For this family the energy functional (9) is
$$
E_{\lambda} = \lambda^{-1} I_1 + \lambda\; I_2 \eqno(20)
$$
where
$$
I_1 =  \int_{0}^\infty \left( \frac{r^2}{2} AN  \phi{}'{}^2
+ A' m \right)dr \;,                                 \eqno(21)
$$
$$
I_2 = \int_{0}^\infty Ae^{-2\phi} \left[ Nw'{}^2 + \frac{(1 -
w^2)^2}{2r^2} \right] dr \;.                              \eqno(22)
$$
Since the energy functonal is extremized at $\lambda=1$, it follows from
(20) that on shell
$$
I_1 = I_2 \;. \eqno(23)
$$
Integrating Eq.(10) one gets $D=-2 I_2$, and therefore
Eq.(23) yields
$$
m(\infty) + D = 0 \;.  \eqno(24)
$$
Thus the total mass can be read off from the monopole term in the
asymptotic expansion (17b) of the dilaton field. The identity (24) is
also true for the limiting solution (15).

\section{Stability analysis}
In this Section I consider
 the time evolution of linear perturbations about the static
solutions described above. I will assume that the time-dependent solutions
remain spherically symmetric and the YM field stays within the
ansatz (3). This is sufficient to demonstrate instability
because unstable modes appear already in this class of perturbations.
For radial perturbations the stability analysis is relatively simple,
because the spherically symmetric gravitational field has no dynamical
degrees of freedom and therefore the perturbations of metric coefficients
are determined by the perturbations of matter fields [11]. To see this
 consider the Einstein equations
$$
\dot{\lambda} = -2 \alpha r  (\dot{\phi} \phi{}' + \frac{2}{r^2} \dot{w}
w')   \eqno(25)
$$
and
$$
\nu{}'-\lambda{}' = \alpha r \left[ {\phi{}'}^2 +\frac{2}{r^2}{w'}^2
+ e^{-2 \nu} ({\dot{\phi}}^2 + \frac{2}{r^2}{\dot{w}}^2) \right]
\;, \eqno(26)
$$
where the functions $\nu$ and $\lambda$ are defined by $e^{\nu}=AN$ and
$e^{\lambda}=N$, and dot denotes the time derivative.
Now, take the perturbed fields $\nu(r)+\delta \nu(r,t)$,
$\lambda(r)+\delta \lambda(r,t)$, $w(r)+\delta w(r,t)$, and
$\phi(r)+\delta \phi(r,t)$, where ($\nu(r), \lambda(r), w(r), \phi(r)$)
is a static non-abelian solution, and insert them into Eqs.(25) and (26).
Linearizing Eq.(25) one obtains
$$
\delta \dot{\lambda} = -2 \alpha  r  (\phi{}' \delta \dot{\phi}
+ \frac{2}{r^2} w' \delta \dot{w}) \;, \eqno(27)
$$
which can be integrated to give
$$
\delta \lambda = -2 \alpha  r (\phi{}' \delta \phi + \frac{2}{r^2}
w' \delta w) \;. \eqno(28)
$$
Linearization of Eq.(26) yields
$$
\delta \nu{}' - \delta \lambda{}' = 2 \alpha r (\phi{}' \delta \phi{}'
+ \frac{2}{r^2} w' \delta w') \;. \eqno(29)
$$
and therefore using Eq.(28) one gets
$$
\delta \nu{}' = -2 \alpha \left[(r\phi{}''+\phi{}')\delta \phi +
\frac{2}{r} (w''-\frac{1}{r} w') \delta w \right] \;. \eqno(30)
$$
The spherically symmetric evolution equations for the matter fields are
$$
-(e^{-\nu - 2 \phi} \dot{w})\dot{} + (e^{\nu -2  \phi} w')'
+ \frac{1}{r^2} e^{\nu - \lambda - 2 \phi}w(1 - w^2) = 0 \;,   \eqno(31)
$$
and
$$
-r^2(e^{-\nu} \dot{\phi})\dot{} + (r^2 e^{\nu}\phi{}')' +
 2 e^{\nu-\lambda- 2 \phi} \left[e^{\lambda} {w'}^2
 +\frac{(1-w^2)^2}{2r^2} \right] = 0 \;.
\eqno(32)
$$
Multiplying Eqs.(31) and (32) by $e^{-\nu}$, linearizing and assuming
harmonic time dependence for the perturbations,
$\delta w(r,t)=e^{i\sigma t} \xi (r)$ and
$\delta \phi (r,t)=e^{i\sigma t} \psi (r)$, one obtains an
eigenvalue problem
$$
-\xi{}'' + (2 \phi{}'- \nu{}')\xi{}' + 2 w' \psi{}' +
2 \alpha w' \left[(r \phi{}''+\phi{}')\psi + \frac{2}{r}(w''-\frac{1}{r}w')
\xi \right]- \frac{1}{r^2} e^{-\lambda} (1-3 w^2) \xi
$$
$$
+ \frac{2 \alpha}{r} e^{-\lambda} w(1-w^2) (\phi{}' \psi+\frac{2}{r^2}
w' \xi)
= {\sigma}^2 e^{-2 \nu} \xi \;,     \eqno(33)
$$
\vspace{0.1cm}
$$
-(r^2 \psi{}')'
 - 4 e^{-2\phi} \left[ w'\xi{}'
- \frac{1}{r^2} e^{-\lambda} w(1-w^2) \xi
- ( {w'}^2 + e^{-\lambda}\frac{(1-w^2)^2}{2r^2} ) \psi
+ \alpha e^{-\lambda} \frac{(1-w^2)^2}{4 r}(\phi{}' \psi + \frac{2}{r^2}
w' \xi) \right]
$$
$$
- r^2 \nu{}' \psi{}'+ 2 \alpha r^2 \phi{}' \left[(r \phi{}''+\phi{}')\psi
+ \frac{2}{r}(w''-\frac{1}{r}w')\xi \right] =  {\sigma}^2 r^2 e^{-2\nu}\psi
\;.   \eqno(34)
$$
If the perturbations satisfy the
boundary conditions $\xi(0)=0$, $\psi(0)=const$,
$\xi(\infty)=0$, and $\psi(\infty)=0$, then the above system is self-adjoint,
hence eigenvalues
${\sigma}^2$ are real. Instability manifests itself in the
presence of at least one negative eigenvalue.

I have used the generalized rule of nodes (Jacobi criterion) to find that
for all values of $\alpha$ the $n$th solution has exactly $n$
negative eigenvalues (I have checked
this up to $n=4$). This is consistent with the fact that the limiting
solution (15) has infinitely many unstable modes which can be seen
easily by inserting (15) into Eq.(33) and considering
perturbations with $\psi=0$. The result that the $n$th solution
has $n$ unstable modes is essential for the Sudarsky-Wald argument [3].
However, by considering a restricted class
of perturbations some directions of instability might have been suppressed.

{\bf Acknowledgement} I am grateful to Peter Aichelburg for stimulating
 discussions.

\section*{Figure captions}
\begin{description}
\item[Fig.1] The YM potential $w_n$ for $n=1, 3, 5$.
\item[Fig.2] The dilaton $\phi_n$ for $n=1, 3, 5$.
\item[Fig.3] The metric coefficient $N_n$ for $n=1, 3, 5$.
\item[Fig.4] $w_1$ for $\alpha=0.01$ (dashed line), $\alpha=1$ (solid line)
and $\alpha=100$ (dotted line).
\item[Fig.5] $N_1$ for $\alpha=0.01$ (dashed line), $\alpha=1$ (solid line)
and $\alpha=100$ (dotted line).
\item[Fig.6] The total mass $\tilde M_n$ (in units $1/e\sqrt{G+a^2}$) as
a function of $\beta=\alpha/(1+\alpha)$.
\end{description}

\end{document}